\documentclass[twoside]{article} 

\makeatletter

\usepackage{amssymb}

\renewcommand\subsection{\@startsection{subsection}{2}{\z@}%
                                     {-3.25ex\@plus -1ex \@minus -.2ex}%
                                     {1.5ex \@plus .2ex}%
                                     {\normalfont\large\bfseries%
                                       \rule{0.6em}{0.6em}\hspace{0.5em}}}
\renewcommand\subsubsection{\@startsection{subsubsection}{3}{\z@}%
                                     {-3.25ex\@plus -1ex \@minus -.2ex}%
                                     {1.5ex \@plus .2ex}%
                                     {
                                        \rule{0.3em}{0.3em}\hspace{0.2em}\rule{0.3em}{0.3em}
                                        \normalfont\normalsize\bfseries
                                        \hspace{0.01em}}}

\@addtoreset{equation}{section}

\usepackage{fancyheadings}
\renewcommand{\sectionmark}[1]%
                {\markboth{\thesection\ #1}{\thesection\ #1}}

\long\def\@makecaption#1#2{
 \vskip 10pt
 \setbox\@tempboxa\hbox{{\small\bf#1:} \small\sl#2}
 \ifdim \wd\@tempboxa >\hsize {\small\bf#1:} \small\sl#2\par
 \else \hbox to\hsize{\hfil\box\@tempboxa\hfil}
 \fi}

\newlength{\halftextwidth}

\RequirePackage{fancybox}

{\vglue 0.2cm \end{minipage}\end{Sbox}%
\shadowbox{\TheSbox}\vglue 0.2cm }

\newcommand{\id}{\mathalpha{\mathbf{1}}}          

\def\square{    {\kern1pt\vbox{
        \hrule height 1.2pt\hbox{
        \vrule width 1.2pt\hskip 3pt
                \vbox{\vskip 6pt}
        \hskip 3pt\vrule width 0.6pt}
        \hrule height 0.6pt}
                \kern1pt}}      

\newcommand{\centeron}[2]{{%
   \setbox0=\hbox{#1}\setbox1=\hbox{#2}%
   \ifdim\wd1>\wd0\kern.5\wd1\kern-.5\wd0\fi
   \copy0\kern-.5\wd0\kern-.5\wd1\copy1%
   \ifdim\wd0>\wd1\kern.5\wd0\kern-.5\wd1\fi
}}

\def\donotprint#1{{}}           

\makeatother

\usepackage{epsf,floatfig}
\begin{document}
\initfloatingfigs

\thispagestyle{empty}
\hfill
\begin{tabular}{l}
HEP-MINN-96-1429 \\
NUC-MINN-96/10-T \\
HEP-MINN-96/08   \\
hep-ph/9606337   \\
\end{tabular}

\vspace{3cm}

\begin{center}
\LARGE \bf The Intrinsic Glue Distribution at Very Small x 
\end{center}

\begin{center}
        {\small Jamal Jalilian-Marian, Alex Kovner, Larry McLerran 
and Heribert Weigert} \\
        {\small\it School of Physics and Astronomy,} \\
        {\small \it University of Minnesota, Minneapolis, MN 55455} 
\end{center}

\vspace{2em}

\begin{abstract}
  We compute the distribution functions for gluons at very small x and
  not too large values of transverse momenta.  We extend the
  McLerran-Venugopalan model by using renormalization group methods to
  integrate out effects due to those gluons which generate an
  effective classical charge density for Weizs\"acker-Williams fields.
  We argue that this model can be extended from the description of
  nuclei at small x to the description of hadrons at yet smaller
  values of x.  This generates a Lipatov like enhancement for the
  intrinsic gluon distribution function and a non-trivial transverse
  momentum dependence as well.  We estimate the transverse momentum
  dependence for the distribution functions, and show how the issue of
  unitarity is resolved in lepton-nucleus interactions.
\end{abstract}

\vfill
\pagebreak

\pagenumbering{arabic}

\section{Introduction}
 
The problem of computing the distribution functions for gluons at very
small x is an old one~\cite{lipatov}. The gluon (and quark)
distribution functions are computable in perturbation theory at large
values of x, but at small x one encounters a so called Lipatov
enhancement.  The precise computation of this enhancement is subject
to much uncertainty primarily because at some point in the evolution
the density of gluons becomes so large that there are mutual
interactions of the gluons which are ignored in the BFKL equation.  In
addition, the behavior at small x also involves knowing the
distribution function at small $Q^2$, and again non-perturbative
information seems to be needed.

Recently a different framework was advocated for the computation of
the gluon distribution functions \cite{mclerran}. The starting point of
this approach is to view a hadron not as a collection of a small
number of partons, but rather as a system with finite parton density.
In the high density situation the natural way to describe the soft
gluons is not as quasi-free particles, but as classical fields with
large amplitude. These classical fields are generated by classical
color charges which represent the valence partons. Once the high
density effects are resummed into the classical fields one may apply
weak coupling methods to calculate quantum corrections.  In this
approach, high gluon densities which prove so problematic in the BFKL
context are a prerequisite for the description of the parton content
of a nucleus wave function via classical gluon fields.

A consistent separation in field and particle like degrees of freedom
can be performed most easily in the infinite momentum frame.  The
object of computation is the intrinsic $x$ and $p_\perp$ contributions
to the infinite momentum hadronic wavefunction in the light cone
gauge.  Distribution functions are given as
\begin{eqnarray}
        G[x,Q^2]  = \int_0^{Q^2}\hspace{-0.2em}
        d^2p_\perp {{dN} \over {dxd^2p_\perp}}
\end{eqnarray}
in terms of the intrinsic parton distributions as computed by taking
the expectation value of the number operator in the state of interest.

In the infinite momentum frame the valence partons are strongly
Lorentz contracted.  If we then look for spread out gluon fields at $x
\ll A^{-1/3}$ interactions between valence partons and soft gluon
fields eikonalize which indeed allows us to take the particle limit
for the fast moving partons.  Here A is the baryon number of a
nucleus.  For a hadron, $x \ll 1$.  The hadron will indeed appear as
an infinitesimally thin sheet on the scale of wavelengths associated
with the momentum fraction $x$.  (We will later see that we will have
to regularize this source by giving it a large but finite momentum and
a longitudinal extent of order $R/\gamma$ where $R$ is its size in the
rest frame and $\gamma$ is its Lorentz gamma factor.  We will find
that nothing in leading order of our computations depends upon the
details of this regularization.) In addition, for a thick nucleus,
since the number of sources of charge per unit area scales as
$A^{1/3}$, we may view the valence partons (quarks) as classical color
charges.  Therefore, somewhat paradoxically the simplest problem to
start with is computing the gluon distributions for a very large
nucleus.
  
We will find later that at very small x, the glue as well as valence
quarks contribute to the charge density seen by a gluon.  The gluons
which contribute to the charge density are all gluons with an x larger
than the x of the gluon whose structure function is being measured.
Therefore the consideration discussed above for nuclei will apply to
hadrons when at sufficiently small x so that the number of gluons at
larger values of x is large.  The advantage of nuclei is that large
densities of charge are generated at larger values of x, and therefore
lower energy per nucleon, than is the case for a single hadron.

A solution of this problem would be useful in a variety of contexts.
The approach we advocate involves knowledge of the nuclear wave
function and is somewhat related to the approach of Mueller for heavy
quarkonia~\cite{mueller}. Our approach in principle allows the
resolution of various phenomenological problems which arise in the
parton cascade model of particle production in heavy ion
collisions~\cite{geiger}. These models provide the initial condition
for hydrodynamic calculations~\cite{bj}. A model which builds in the
space-time structure we advocate and uses the information we have
generated for the infinite momentum frame wave functions is given in
Ref.~\cite{kovner}.

The theory which results at the classical level is basically a
Yang--Mills theory in the presence of a source
\begin{eqnarray}
      J^+_a = \delta(x^-) \rho_a (x_\perp)
\end{eqnarray}
The measure which generates the expectation values of gluon fields,
corresponding to distribution functions is
\begin{eqnarray}
     \int [dA] [d\rho ]
 ~\exp\left(-\int d^2x_\perp{1 \over {\mu^2}} Tr\rho^2(x_\perp) 
 \right)
 ~\exp{iS}
\label{MV}
\end{eqnarray}
where $S$ is the ordinary gluon action in the presence of the external
current $J$.  The parameter $\mu^2$ is the valence color charge per
unit area (scaled by a factor $1/(N_c^2-1)$).  In leading order, the
expectation value is given by a classical field which is a solution of
the Yang-Mills equation\footnote{For conventions on the use of the
  coupling constant please see the next section.}
\begin{eqnarray}
     D_\mu F^{\mu \nu} = g^2 J^\nu
\end{eqnarray}
which is then averaged over different values of $\rho$

In the limit where the gluon field generated by these valence quarks
is treated classically, the gluon field is a non-Abelian
Weizs\"acker-Williams field, and has the form
\begin{eqnarray}
     A^+ = A^- = 0
\end{eqnarray}
and
\begin{eqnarray}
     A^i = \theta (x^-) \alpha^i (x_\perp)
\end{eqnarray}
The field $\alpha^iS$ is a two dimensional ``pure gauge''
\begin{eqnarray}
     \alpha^i = -{1 \over {i}} U \nabla^i U^\dagger
\end{eqnarray}

The physical justification for the non-Abelian Weizs\"acker-Williams
field is that because the source of charge is confined to a thin
sheet, the solution must solve the free equations of motion everywhere
but on the sheet.  The solution is therefore a gauge transform of zero
field on either side of the sheet.  The discontinuity of the fields
across the sheet gives the charge density.

It was suggested by McLerran and Venugopalan that this simple model
should give a decent approximation for the soft glue distribution
function.  It turns out, however that the corrections to the
distribution function calculated in this way are large at small $x$.
Technically there are two sources for these corrections, although both
have the same physical origin.

First, as we will show the behavior of the correlation function
calculated in this simple minded approach is singular at small
$p_\perp$.  The flaw in the treatment of Ref.~\cite{mclerran} was that
the source of charge was not treated as an extended distribution which
tends to a delta function only in the infinite momentum
limit~\cite{rajiv,yuri}. Physically it is clear that the charge
density is indeed spread out on the scale of the characteristic
longitudinal momentum of the hard particles which generate this
density.

The second source of large corrections is basically the same small $x$
enhancement as in standard perturbative calculations. As was shown
in~\cite{ayala} the quantum corrections to the distribution functions
calculated in terms of the classical fields become large at small $x$,
the enhancement factor being the infamous $\alpha \log 1/x$.

The main goal of this paper is to show that the two corrections are
physically related and to outline a solution to both problems.  We
will show that the small $x$ enhancement arises from quantum
fluctuations with large longitudinal momentum.  We show that such
configurations may be successively integrated out by using
renormalization group techniques reminiscent of the Wilson block spin
method. This approach can also be interpreted in terms of the
adiabatic or Born-Oppenheimer approximation extensively used in atomic
physics.  Integrating out hard quantum fluctuations is equivalent to
including the harder gluons into what we call the charge density
$\rho$ in Eq.~(\ref{MV}) while calculating the distribution of the
softer glue.  It therefore leads to ``renormalization'' of the charge
density and endows it with nontrivial and calculable longitudinal
structure.

This modified, momentum dependent distribution of source strengths
leads to infrared non-singular correlation functions.  We argue that
the result is sensitive only to the average charge squared per unit
rapidity per unit transverse area of the source.

The outline of this paper is as follows:

In section \ref{sec:extstruct} we study the classical problem of
computing the fields associated with a source of charge which is
extended in $x^-$.  We find the general solution to this problem in
light cone gauge.  We compute the resulting distribution functions
assuming that the source is random in $x_\perp$ and $x^-$, but with a
weight of charge squared per unit rapidity per unit area which is
specified.

In section \ref{sec:rgcharge}, we show by using the renormalization
group techniques how to generate classical fields at some rapidity
scale $y$.  This involves perturbatively integrating out modes at
larger values of rapidity (smaller values of $x^-$).  This integration
generates an effective Lagrangian which has a self similar form,
namely that at each step of the procedure it is similar to the
McLerran-Venugopalan model but with a charge per unit area which is
rapidity dependent.  We show that this effective theory is equivalent
to that discussed in section \ref{sec:extstruct}.  We derive the
renormalization group equations for the charge squared per unit area
per unit rapidity as measured at some transverse momentum scale $Q^2$
and rapidity $y = y_0 + \ln(x^-_0/x^-)$, where $y_0$ is the nucleus
rapidity, and $x^-_0 \sim R/\gamma$. The calculations in this section
rely on several simplifying approximations, which we discuss.
 
In section \ref{sec:rgchi}, we study the renormalization group equations
for the charge squared per unit rapidity per unit area as a function
of $Q^2$ and $y$.  This equation is closely related to the usual
evolution equations for distribution function which appear in standard
perturbative treatments. It can be viewed as a nonlinear version of
DGLAP equation. We show that for $p_\perp$ much larger than the
momentum scale associated with the charge squared per unit rapidity
integrated over all rapidities larger than that at which we measure
the structure functions (which we will refer to as $\chi (y,Q^2)$),
the nonlinearities in the RG equation become unimportant. In this
regime the equation basically describes the double log DGLAP
evolution.  At lower momenta our equation can be thought of as a
nonlinear variant of the BFKL equation, although to make the relation
precise one would have to consider some virtual corrections in
addition to those accounted for in our derivation.  At low transverse
momentum, we show that the evolution equation saturates.  We discuss
the consistency issues which are necessary for a solution of this
equation within the set of approximations for which our derivation is
valid.

In the final sections we summarize our results.  We show how our
results are consistent with unitarity in deep inelastic scattering.
We estimate the total cross section at fixed $Q^2$ as $x$ approaches
zero.  We argue that a computation of the charge squared per unit
rapidity per unit area would allow a computation of the total
multiplicity in hadronic interactions.  We discuss the possible
universality of our results and their possible generalization to the
description of nucleons at small x.  We discuss some of the many
problems which are not yet solved within the approach advocated in
this paper.

\section{Modification of the Source Strength Due to Extended 
  Structure in $x^-$}\label{sec:extstruct}

In this paper we will use gauge potentials scaled such that the
covariant derivative reads $D_\mu[A] = \partial_\mu -i A_\mu$. The
classical gluonic action is of the form $-\frac{1}{4 g^2} F^2$ and
hence a $g^2 J^\mu$ term in the classical equations of motion. In this
setup gauge transformations $U$ are most economically parameterized
via $U(x) = \exp\left[ i \Lambda(x)\right]$ transforming $A$ as $A \to
U\left[ A - \frac{1}{i}\partial\right] U^{-1}$.  We will be also using
matrix notation, e.g.  $\rho=\rho^a T^a$, where $T^a$ are the
normalized hermitian generators of the $SU(N_c)$ group in the
fundamental representation, $2{\mathrm{Tr}}\, T^aT^b=\delta^{ab}$.

In the original McLerran-Venugopalan approach~\cite{mclerran}, the
source strength was assumed to have the form
\begin{eqnarray}
     J^+_a (x^-,x_\perp) = \delta (x^-) \rho_a(x_\perp)
\end{eqnarray}
and to be distributed with the Gaussian
weight
\begin{eqnarray}
     \int [d\rho ] 
     \exp\left[ -{1 \over {\mu^2}} \int d^2x_\perp 
       {\mathrm{Tr}}\,\rho^2(x_\perp) \right]
\end{eqnarray}
where $\mu^2$ is the charge per unit area.

The solutions to the Yang Mills equation in $A^+ = 0$ gauge have
vanishing $A^-$. Their transverse components $A^i$ are determined
through
\begin{eqnarray}
     \nabla_i \partial^+ A^i +[ A_i , \partial^+ A^i] = g^2 J^+
\label{eom}
\end{eqnarray}
together with
\begin{equation}
F^{ij}=0
\end{equation}
It was argued that the solution was of the form
\begin{eqnarray}
     A^i (x) = \theta (x^-) \alpha^i (x_\perp)
\end{eqnarray}
In this solution, the commutator term in Eq.~(\ref{eom}) was ignored
since it involves the commutator of the field at the same point in
$x^-$.

Ignoring the commutator term is however not justified.  It is clear
that this term in Eq.(\ref{eom}) is very singular and involves a
product of $\delta (x^-)$ and $\theta (x^-)$. To make sense of this
structure, we must understand the evolution of the field across the
delta function singularity.  This can only be done if we know the
structure of the source in $x^-$.  In fact, as shown in~\cite{rajiv}
ignoring this problem leads to infrared singular distributions.

Let us introduce the space-time rapidity variable 
\begin{eqnarray} 
  \label{eq:rapidity}
  y = y_0+\ln (x^-_0/x^-) 
\end{eqnarray}
which will be useful for $x^- > 0$.  We will assume that the source
strength is non-vanishing only for positive $x^-$, and we will work in
a gauge where the fields $A^i$ vanish for $x^- < 0$.  The rapidity
$y_0$ is the momentum space rapidity of the nucleus, and the parameter
$x^-_0$ is the typical Lorentz contracted size of the nucleus $x^-_0
\sim R/\gamma$.

In the next section, we will use renormalization group arguments to
show that there is a non-trivial induced source strength extending
beyond the volume occupied by valence partons which is driven by gluon
modes at longitudinal momentum larger than that at which we measure
the gluon distribution. This is parameterized by some strength of
charge squared per unit area per unit space-time rapidity $\mu^2(y,
Q^2)$, and by the charge per unit area at rapidities greater than $y$
\begin{eqnarray}
  \chi(y,Q^2) := \int_y^{\infty} \hspace{-.5em}
                 dy^\prime \mu^2 (y^\prime, Q^2)
\label{chi}
\end{eqnarray}
The parameter $Q^2$ appears because we must specify at what value of
$Q^2$ we are measuring the distribution function.  It has precisely
the same meaning as in perturbative QCD calculations, namely the
transverse scale at which a parton is resolved~\cite{levin}.  It
should not be confused with the intrinsic transverse momenta of the
fields.

Accounting for the space-time rapidity dependence of the source
strength, we therefore are lead to consider the distribution
\begin{eqnarray} \int [d\rho ]~ \exp \left[ -
\int\limits_0^{\infty} \hspace{-0.2em} dy \int\hspace{-0.2em}
d^2\!x_\perp {  {\mathrm{Tr}}\,
 \rho^2(y,x_\perp)\over {\mu^2(y,Q^2)}} \right]
\label{rhoint}
\end{eqnarray} 
In this equation, $\rho$ is the charge density per unit transverse
area per unit space-time rapidity\footnote{The parameter $\mu^2(y)$
  controls the magnitude of the fluctuations of the charge density at
  fixed rapidity $y$. Since there is no charge density at rapidities
  greater than the rapidity of the nucleus $y_0$, the function
  $\mu^2(y)$ should vanish for $y\ge y_0$. The rapidity integrals in
  Eqs.~(\ref{chi}), (\ref{rhoint}) are therefore effectively cutoff at
  this upper limit.}.  In the previous work we took great pain to
argue that the charge could be treated classically on transverse
scales which are large compared to the density of partons per unit
area.  This was because on this scale there is a large number of
partons contributing to the source, and therefore the charges were in
a large dimensional representation of the color group.  This allowed a
classical treatment.

The longitudinal structure is a new ingredient. Why can we still
approximate the partons (gluons) that couple to soft glue by a
classical source?  The physical reason is easy to understand: for
these high momentum gluons the coupling is weak, so that to change the
field (the soft glue) by a correction of order one, one must have many
(hard) gluons contributing.  In the next section, we will see that the
induced source of gluons is in fact slowly varying in rapidity
\begin{eqnarray} 
{{d^2 \rho} \over {dy^2}} \sim \alpha {{d\rho} \over {dy}} 
\end{eqnarray}
Again, we will see this justified in more detail in the next section.

Therefore these sources of charge come from an extended region of
space-time rapidity with a typical contribution at a rapidity far
greater than that of the field we are computing.  The source will
therefore appear to be infinitesimally thin in the variable $x^-$.

We must solve Eq.~(\ref{eom}) in the presence of source with a
prescribed rapidity dependence.  This is best done in terms of the
rapidity variable $y$ introduced in (\ref{eq:rapidity}).
Eq.~(\ref{eom}) becomes
\begin{eqnarray} 
D_i {d \over {dy}} A^i = g^2 \rho (y,x_\perp) 
\end{eqnarray} 
In this equation, because of the extended structure in rapidity, the
term which involves the group cross product of two $A^i$ fields cannot
be ignored.

The formal solution to this equation can be found by introducing the
line ordered phase
\begin{eqnarray}
     U(y,x_\perp) = U_{\infty,y}(x_\perp) = \hat{\mathsf P}
     \exp\left[i \int^\infty_y dy^\prime \Lambda(y^\prime,
     x_\perp)\right] \label{V}
\end{eqnarray}
representing a parallel transport operator along a straight line at
fixed $x^+$ and $x_\perp$ connecting $y$ to $\infty$ ($x^- = 0$).
Recall, that due to vanishing of the transverse magnetic field
($F^{ij}=0$), the vector potential should be a ``two dimensional pure
gauge''. We let therefore
\begin{eqnarray}
     A^i(y,x_\perp) = i U \nabla^i U^{-1}
\end{eqnarray}
This leads to the equation for $\Lambda$
\begin{eqnarray}
     \nabla^2 \Lambda =  -g^2U^{-1} \rho\  U
\end{eqnarray}

The above equation may be solved directly numerically.  Imagine we
have a grid in rapidity $y$ and transverse coordinates.  We define the
lattice spacing in rapidity as $a_y$.  The above equation can be
written as
\begin{eqnarray}
\lefteqn{     \nabla^2 \Lambda(y,x_\perp)} 
\\ && \nonumber 
= -g^2 
\left(\hat {\mathsf P} \exp\left[i\int^\infty_{y+a_y}
    \hspace{-1em}
     dy^\prime\Lambda(y^\prime,x_\perp)\right]\right)^{-1}
\hspace{-.8em} \rho(y,x_\perp) 
\left(\hat {\mathsf P} \exp\left[i\int^\infty_{y+a_y}
    \hspace{-1em}
    dy^\prime \Lambda(y^\prime,x_\perp)\right]\right) \label{lambda}
\end{eqnarray}
The solution at $y$ depends only upon the function $\Lambda $ at
larger values of rapidity.  This equation may therefore be solved
iteratively starting at some maximum $y_{\mathrm{max}}$ beyond which
the source vanishes.

It turns out, however that we do not need to know an explicit solution
in order to calculate the distribution function. For that we have to
perform the integration over the source strength $\rho (y,x_\perp)$.
Let us change the variables in the path integral Eq.~(\ref{rhoint})
from $\rho(y,x_\perp)$ to $\Lambda(y,x_\perp)$.  The Jacobian of this
transformation does not depend on $\rho$.  This is easily verified
noting that the Jacobi matrix is triangular in $y$ and due to the $y$
orderings involved in the relation between $\rho$ and $\Lambda$ has no
interactions on the diagonal.  We find therefore that for any function
$O(\rho)$
\begin{eqnarray}
\lefteqn{\int [d\rho ]~ \exp \left[ -
\int\limits_0^{\infty} \hspace{-0.2em} dy \int\hspace{-0.2em}
d^2\!x_\perp { {\mathrm{Tr}}\,
 \rho(y,x_\perp)^2 \over {\mu^2(y,Q^2)}} \right]O(\rho)}   
\\ \nonumber &= &
 \int [d\Lambda ]~ \exp \left[ -
\int\limits_0^{\infty} \hspace{-0.2em} dy \int\hspace{-0.2em}
d^2\!x_\perp { {\mathrm{Tr}}\, 
  \left(\nabla^2 \Lambda(y,x_\perp)\right)^2 \over 
  {g^4\mu^2(y,Q^2)}} \right]O(\Lambda)
\label{lambda1}
\end{eqnarray} 
Since the classical fields $A_i$ are given as explicit functions of
$\Lambda$, and our aim is to compute the distribution function
\begin{eqnarray}
    G_{ij}(y,x_\perp;y^\prime,x_\perp^\prime) = 
\langle A_i(y,x_\perp) A_j(y^\prime,x_\perp^\prime)\rangle
\end{eqnarray}
this form of the path integral is very convenient.

 We first note
that
\begin{eqnarray}
     A^i(y,x_\perp) = 
     \int\limits^\infty_y \hspace{-0.2em} dy^\prime \
     U_{\infty,y^\prime}(x_\perp) 
     \left( \nabla^i \Lambda(y^\prime,x_\perp) \right)
     U_{y^\prime,\infty}(x_\perp) 
\end{eqnarray}
Now we perform the integrations over $\Lambda $ by expanding the path
ordered phases.  It is most conveniently done by expanding the
exponentials to first order on the rapidity grid with grid spacing
$a_y$.  This is a valid procedure as long as the function $\mu^2$ is
regular in the sense that $\lim\limits_{a_y\rightarrow
  0}a_y\mu^2(y)=0$. We then perform all possible contractions with the
propagator corresponding to the Gaussian weight in the path integral
over $\Lambda$.  Let us group together terms of the same order in the
coupling constant.  In zeroth order we have
\begin{eqnarray}
G_{ij;ab}^0(y,x_\perp;y^\prime,x_\perp^\prime ) = g^4 \delta_{ab}
     \int_{{\mathrm{max}}_{y,y'}}^\infty \hspace{-1.5em} dy \
      \mu^2(y,Q^2) \nabla_i \nabla_j^\prime 
      {1 \over {\nabla^4}}(x_\perp,x_\perp^\prime)
\end{eqnarray}
Some comments are in order concerning the inversion of the operator
$\nabla^4$, since there is an infrared singularity in the inversion.
Recall that the sources of interest ultimately arise from individual
nucleons.  Therefore all effects of sources die off at transverse size
scales larger than $1/\Lambda_\mathrm{QCD}$.  The charge itself
averaged over such transverse size scales also vanishes.  This means
that the Green's function should be defined with boundary conditions
that ensure its vanishing at distance $1/\Lambda_\mathrm{QCD}$.  In
other words, whenever an infrared cutoff is needed for proper
definition of an inverse of a differential operator, it should be
taken of the order of $1/\Lambda_\mathrm{QCD}$.  We will see that the
quantities of physical interest are only very weakly dependent upon
this non-perturbative length scale, but nevertheless such dependence
does not disappear entirely.

Here and in all that follows, we will define 
\begin{eqnarray}
     \gamma (x_\perp) & := & {1 \over {\nabla^4}}(x_\perp) 
                 = {1 \over {8\pi}} x_\perp^2 
                       \ln (x_\perp^2\Lambda^2_\mathrm{QCD}) + 
                       \gamma(0)
                        \label{k4reg}
\end{eqnarray}
where $\gamma(0)$ denotes an (infrared divergent) constant which
ensures the vanishing of this Green's function as $x_\perp$ approaches
the infrared cutoff $1/\Lambda_\mathrm{QCD}$.  Fortunately, the
correlation function we will calculate below does not depend on the
value of $\gamma(0)$ and the only infrared sensitivity that remains is
through the logarithmic term in Eq.~(\ref{k4reg}).

In the first order, a quick computation gives
\begin{eqnarray}
 G_{ij}^1(y,x_\perp;y^\prime,x_\perp^\prime ) & = &
 -{1 \over {2\!}}  \delta_{ab} (-N_c)
\left[g^4 \int_{{\mathrm{max}}_{y,y'}}^{\infty} \hspace{-1.5em} 
  dy^{\prime\prime} \mu^2(y^{\prime\prime},Q^2)\right]^2 \nonumber \\
& & \left( \nabla_i \nabla_j^\prime 
\gamma(x_\perp -x_\perp^\prime)\right) 
[\gamma (x_\perp -x^\prime_\perp)-\gamma(0)]
\end{eqnarray}
In this equation, $N_c$ is the number of colors.

Similarly, in $n'th$ order, we find
\begin{eqnarray}
G_{ij;ab}^n(y,x_\perp;y^\prime,x_\perp^\prime ) & = & 
     (-1)^n\delta_{ab}
     { (-N_c)^n\over {(n+1)!}}
\left[ g^4\int_{{\mathrm{max}}_{y,y'}}^\infty \hspace{-1.5em} 
  dy^{\prime\prime} \mu^2(y^{\prime\prime},Q^2)\right]^{(n+1)} 
\nonumber \\ & & 
  \nabla_i \nabla_j^\prime 
      \gamma(x_\perp-x_\perp^\prime) 
    \left\{ \gamma(x_\perp-x^\prime_\perp)-\gamma(0)\right\}^n 
\end{eqnarray}
The tadpole terms which take care of subtractions of $\gamma(0)$
appear through the normal ordering of the path ordered exponential.
This calculation can be found in Appendix A.

We can now sum the series and find a representation for the
correlation function as (assuming $y > y^\prime$)
\begin{eqnarray}
     G_{ij}^{ab}(y,x_\perp;y^\prime,x_\perp^\prime) & = &  
 -\delta^{ab}\left( \nabla_i \nabla_j^\prime 
\gamma(x_\perp-x_\perp^\prime) \right)
 {1 \over { N_c\left[ 
  \gamma (x_\perp-x^\prime_\perp)-\gamma(0)\right]}} 
\nonumber \\ && 
  \left(1 - \exp\left\{g^4N_c \chi(y,Q^2) 
[\gamma(x_\perp-x^\prime_\perp)-\gamma(0)]\right\} \right)
\end{eqnarray}
where $\chi(y,Q^2)$ is the total charge squared per unit area at
rapidity greater than the rapidity $y$ introduced in~(\ref{chi}).

Finally, for the distribution function we get
\begin{equation}
G_{ii}^{aa}=\frac{4(N_c^2-1)}{N_c x_\perp^2}
     \left[1-\left(x_\perp^2\Lambda_\mathrm{QCD}^2\right)^{
     \frac{g^4N_c}
      {8\pi}\chi(y)x_\perp^2}\right]
\label{propagator}
\end{equation}

This correlation function has an amusing structure.  At small
transverse distances where $\gamma $ approaches zero, the correlation
function tends to the perturbative correlation function, that is its
value in lowest order in an expansion in $g^2$.  At short distances,
the theory is perturbative.
\begin{figure}[hbt]
\epsfysize=7cm
\begin{center}
\begin{minipage}{4cm}
\epsfbox{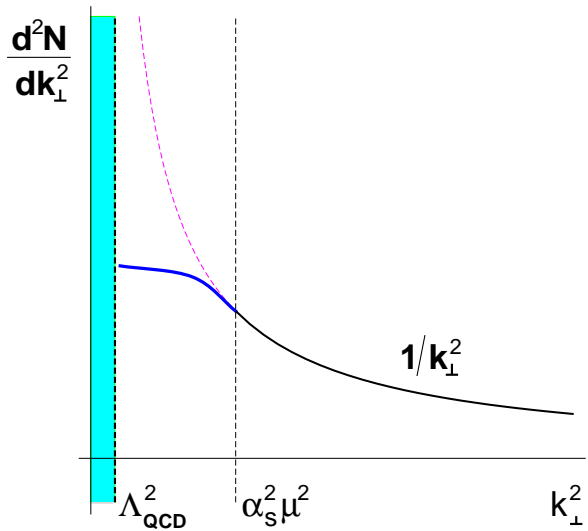}
\end{minipage}
\end{center}
\caption{\label{fig:softening} \small \itshape 
  The distribution at fixed $x$ as a
  function of intrinsic transverse momentum $k_\perp^2$. We obtain
  considerable softening at small $k_\perp^2$ compared to the
  perturbative $1/k_\perp^2$ behavior}
\end{figure}

At large transverse distances (but of course still much smaller than
$1/\Lambda_\mathrm{QCD}$), the correlation function dies off like
$1/x^2$.  It's Fourier transform at small momenta therefore behaves as
\begin{eqnarray}
  G(y,y^\prime = y, k_\perp) \sim \ln(k_\perp^2/g^4N_c\chi(y))
\end{eqnarray}
This is in contrast to the behavior at larger transverse momentum
where this correlation functions rises like $1/k_\perp^2$ as $k_\perp$
decreases, in agreement with the perturbative result. The correlation
function is therefore much softened at small $k_\perp$.  This behavior
is shown in Fig.\ref{fig:softening} The characteristic momentum scale
which differentiates between the non-perturbative and perturbative
regions is $ k_\perp^2 \sim g^4N_c\chi(y)$, that is $g^4$ times the
charge per unit area at rapidities greater than the rapidity at which
the correlation function is measured\footnote{As we will see in the
  next section, the contribution of the gluons to $\chi$ is
  proportional to $N_c$ at large $N_c$. In the large $N_c$ limit the
  coupling constant scales as $g^2N_c=const$. The crossover scale
  therefore has the correct large $N_c$ scaling behavior.}.  This
non-perturbative regime is nevertheless a weak coupling regime.  Only
when $k_\perp^2 \sim \Lambda^2_\mathrm{QCD}$ does the coupling become
strong, and weak coupling methods can no longer be used.

It is worth noting that the dependence upon $\Lambda_\mathrm{QCD}$ is
very weak.  At large transverse momenta the Fourier transform of the
distribution function $G$ does not depend on $\Lambda_\mathrm{QCD}$.
At large separations there is saturation, and there is again no
dependence upon $\Lambda_\mathrm{QCD}$. The dependence is really only
in the region of very small momenta ($k_\perp \ll\alpha_s\chi$), where
our approximation is in any case not valid.

This result is almost consistent with the structure which was
advocated by McLerran and Venugopalan~\cite{mclerran}.  They had
argued that at small transverse momentum, the above correlation
function should approach a constant.  It does up to logarithmic
corrections.  (The line of reasoning in~\cite{mclerran} was however
incorrect since it was based on an analysis of an equation that did
not properly handle the induced charge associated with the gluon
field, that is the $[A_i,\partial^+ A^i]$ term in the equation which
determines the gluon field in terms of the external charge density).

\section[Renormalization Group Improved Charge
Distribution]{Renormalization Group Improved Charge Distribution and
  Gluon Field}\label{sec:rgcharge}

In previous work, it was shown that radiative corrections to the
distribution functions computed in the McLerran-Venugopalan model are 
large~\cite{ayala}.  The first order correction comes from the diagram, 
Fig.\ref{fig:twolegs}.

The modification of the distribution function is of order $\alpha_s
\ln (1/x) \ln (k_\perp/\alpha_s \mu)$ at large $k_\perp$ and small
$x$.  For $k_\perp \sim \alpha_s \mu$, the natural $k_\perp$ scale in
the problem, the corrections are of order $\alpha_s \ln (1/x)$.  In
any case, for small values of $x$, these corrections become large and
cannot be ignored.

\begin{floatingfigure}{0.5\textwidth}
\epsfysize=4.5cm
\begin{center}
\begin{minipage}{4cm}
\epsfbox{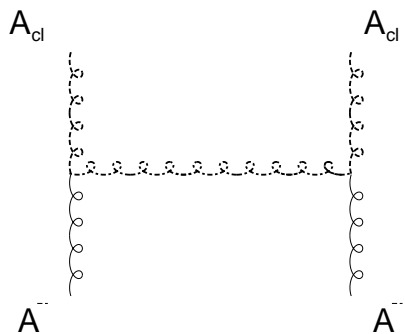}
\end{minipage}
\end{center}
\caption{\label{fig:twolegs} \small \itshape 
  The leading perturbative correction to the gluon distribution beyond
  the classical field approximation.  The fat lines denote gluons with
  large longitudinal momentum $p^+$.  The momentum of the thin lines
  is $k^+$. The large logs come from the kinematical region
  $k^+/p^+\ll 1$.}
\end{floatingfigure}

In this section, we will set up a renormalization group procedure that
sums up these corrections. The method of analysis is the following: We
first consider the bare McLerran--Venugopalan model at a fixed valence
charge density, and a fixed ultraviolet cutoff in the longitudinal
momentum $P^+$.  Physically $P^+$ is of the order of the longitudinal
momentum of the nucleus.  It sets the scale of size for the
longitudinal extent of the nucleus.

The renormalization group is implemented by considering the effective
Lagrangian at a scale of momentum $k^+$ much less than $P^+$ but where
$\alpha_s \ln (1/x) \ll 1$.  To generate this effective Lagrangian, we
integrate out quantum fluctuations with momentum $k^+ \le q^+ \le
P^+$.  This procedure, as will be seen generates a new effective
Lagrangian of the same form as the original one, but with an
additional charge squared per unit area.  The typical scale of
fluctuation of this additional charge squared per unit area is
$\mu^2(y,Q^2) dy$ where $dy = -\ln (x)$ and $Q^2$ is a typical
transverse momentum resolution scale at which the gluon distribution
is ultimately measured.

Since the form of the Lagrangian is unchanged under integration of
these high momentum modes, except for the overall scale factor
$\mu^2$, the procedure can then be repeated and yet lower momentum
modes can be integrated out. Importantly, as long as the coupling
constant is small and also $\alpha_s \ln (x_1/x_2)$ is small the
quantum fluctuations can be integrated out perturbatively, so that the
computation is controlled.  This perturbative treatment considers
fluctuations around the classical solution in the region of the phase
space which is being integrated out, as small. It is important to
realize, however that the classical solution itself in this region of
momenta is changed relative to the one at the previous step of the RG
procedure. This is because it solves classical equations in the
presence of the additional charge $\mu^2(y,Q^2) dy$.  This allows us
to generate an effective Lagrangian at some scale of $x \ll 1$.  This
is in a region where the naive McLerran-Venugopalan model would have
broken down.

Analogously, we can also allow the transverse momentum cutoff $Q^2$ to
be changed independently by a renormalization group transformation.
This corresponds to perturbatively integrating out quantum modes in a
phase space region with transverse momenta between $Q_1$ and $Q_2$.
This latter RG transformation is the counterpart in our approach of
the standard perturbative renormalization group scaling.

In the process of doing these transformations we develop a set of 
renormalization group equations for $\mu^2 (y,Q^2)$.  These equations
determine the rapidity and $Q^2$ dependence of this parameter. 

We also determine the equation for the gluon field.  It will turn out
that this equation is a little more complicated than that in the
previous section, since the induced charge depends upon the gluon
field strength squared at the previous step of the renormalization
group analysis.  We argue that it should be a reasonable approximation
to replace this field strength squared by its average value, in which
case the equations described in the previous section can be derived.
There are corrections to this approximation which are in principle
computable.  It is precisely this approximation which makes the
fluctuations in the charge density uncorrelated in space-time
rapidity.  Inclusion of these corrections will induce correlations.
These correlations will however on the average not contribute to
building up the charge density.

It should be noted that there are other sources of correlation in
longitudinal phase space.  These arise from the classical field itself
which is recomputed at each stage of the renormalization group
analysis.  Although it is true that the source of the color field is
largely uncorrelated, for a given source, there are still long range
correlations built into the color field which would yield non-trivial
multiplicity correlations in rapidity.

Since the process closes under iteration, it is sufficient for us to
show how we integrate the degrees of freedom as the Lagrangian changes
scales in the Nth to the N+1st step of the renormalization grouping.
This is what will be demonstrated below.

We begin our analysis with the McLerran-Venugopalan action
\begin{eqnarray}
     S = i\int d^2x_\perp {1 \over {\chi}} 
         {\mathrm{Tr}}\,\rho^2(x_\perp) 
        -{1\over g^2}\int d^4x {1 \over 4} 
           F_{\mu \nu} F^{\mu \nu} 
        + \int d^4x A^- J^+ 
\end{eqnarray}
where
\begin{eqnarray}
     J^+(x) = \delta (x^-) \rho (x_\perp)
\end{eqnarray}
We are of course working in $A^+ = 0$ gauge.

Now suppose we have a solution to the classical equations of the form
\begin{eqnarray}
      A^+ & = & A^- = 0 \nonumber \\
      A^i & = & \theta (x^-) \alpha^i (x_\perp)
\label{class}
\end{eqnarray} 
     
It is understood in the above expressions that the longitudinal delta
function (as well as theta function) is regularized on the scale
$1/P^+_N$. Here $P^+_N$ is the typical longitudinal momentum of the
fluctuations which have been integrated out in the previous step of
the RG. At the very first step, of course $P^+$ is the typical
momentum associated with the nucleus or hadron.  Clearly, the
knowledge of the precise structure of the charge density on the scales
of order $1/P^+_N$ is necessary to determine the behavior of the
classical solution at these scales.

However, in the following we will only need to know the structure of
the solution at longitudinal momenta much smaller than $P^+_N$. This
is so even though we will integrate over the fluctuations in the
entire momentum range
\begin{eqnarray}
     P^+_{N+1} < k^+ < P^+_N
\end{eqnarray}
where $P^+_{N+1}\ll P^+_N$, but is still large enough so that
$\alpha_s \ln (P^+_N/P^+_{N+1}) \ll 1$ (here $\alpha_s$ is evaluated
at the scale $\chi$ and $\chi$ is assumed to be $\chi \gg
\Lambda_\mathrm{QCD}$).  The reason is that the dependence on the
upper cutoff is only logarithmic, and the bulk of the contribution
comes from much smaller momenta. To leading order therefore the
results do not depend on the precise behavior of the classical field
at the upper cutoff scale. Such is the magic of the logarithm, which
enamored so many field theory practitioners. At momenta far below the
cutoff the classical field does indeed have the structure
$A^i(k)\propto 1/k^+\alpha^i(k_\perp)$, which is equivalent to
Eq.~(\ref{class}).

To compute the effective action, we must integrate out the
fluctuations around the classical solution.  So long as we only
generate an effective action at a scale $k^+$, so that $\alpha_s \ln
(P^+_{N+1}/k^+) \ll 1$, then these fluctuations are small.  We
therefore have three types of fields to consider.  There is the
classical background field, the small fluctuation field at the scale
of interest and the fields at lower momentum scale.  We need only keep
terms in the action which are at most quadratic in the small
fluctuation field.  We will denote these fields in the following
manner:

The field $A_\mathrm{cl}^N$ will be the classical background field.
The label $N$ refers to the $N'th$ step in the renormalization group
procedure.  This classical background field will be modified as the
renormalization group procedure iterates.  We will also write
\begin{eqnarray}
  A_\mathrm{cl}^N = \theta_N(x^-) \alpha^N (x_\perp)  
\end{eqnarray}
In this equation, as mentioned before the step function is a step
function on distance scale larger than that which we have previously
integrated out.

There are the small fluctuations fields at the step $N$ which are
within the momentum range that we integrate out.  We will refer to
these fields as $\delta A^N$

Finally there are the fields which are fluctuations around the
classical solution at momentum scales much less than that where we
perform the integration. These fields are not small.  They are denoted
as $A^N$.

The contribution to the effective action associated with the small
fluctuation field is
\begin{eqnarray}
  & &   \delta S = {1 \over g^2}
\int d^2\!x_\perp dx^+  \left\{ \int
{{dk^+dp^+} \over {4\pi^2}}
  \left(  { 1 \over 2} \delta A^N(k^+) D^{-1}_N(k^+,p^+) \delta A^N(p^+) \right) \right.
 \nonumber +\\
&   &  2f_{abc} \int {{dk^+}\over{2\pi}} \left. \int^{P_N^+}_{-P_N^+} 
{dk^{+\prime} \over {2\pi}}  
\alpha^{iN}_a(x_\perp) A^{-N}_c(k^{+\prime},x^+,x_\perp) 
  \delta A_{ib}^N
(k^+,x^+,x_\perp) \right\} 
\label{deltas}
\end{eqnarray}
In this equation, the momenta $k^+$ and $p^+$ are in the range between
the cutoffs $P_N^+\le |k^+|, |p^+|\le P_{N-1}^+$.  The momentum
$k^{+\prime}$ is typically much softer than the lower cutoff, and
therefore also much softer than $k^+$.  The quantity $D^{-1}_N$ is the
inverse propagator in the background field. It depends on both the
fields $A_\mathrm{cl}^N$ and $A^N$.

We have approximated the linear term in the small fluctuation by
keeping only the eikonal part of the interaction vertex, that is the
coupling between the transverse components of the hard field and the
minus component of the soft field.  This will generate an effective
action with only $+$ components of currents affected by integrating
out the high momentum modes.  The terms we neglected are suppressed by
factors $k^{+\prime}/k^+$ and are sub-leading in the small $x$ region.

It is also understood that the transverse momenta of all the fields in
Eq.~(\ref{deltas}) are bounded from above by some transverse cutoff
$Q$.  This is consistent with both, the BFKL approach, where all
transverse momenta are roughly the same, and the leading log (or
double log at small $x$) Altarelli - Parisi evolution, where the
momenta are bounded by the momentum of the external probe.  By
imposing such a cutoff we restrict ourselves to transverse momenta
which are not parametrically large.  This point can be appreciated by
examining the Feynman diagrams.  Consider for example the diagram that
gives the leading correction to the distribution function beyond the
classical field approximation.  It is depicted in
Fig.\ref{fig:twolegs}.  The corrections of this type with arbitrary
number of insertions of the background field have been calculated
in~\cite{ayala}. For our present purposes it is enough to consider the
classical field expanded to first order in the charge density $\rho$.
The diagram then is precisely the same as that of the standard
perturbation theory.  After the integration over the frequency $k^-$
is performed, the correction to distribution function is proportional
to
\begin{eqnarray}
{1\over{k^+k_\perp^2}}\int {{d^2p_\perp}\over{p_\perp^2}}
\int_{k_N^+}^{k_{N-1}^+}
dp^+{{p^+}\over{\left(p^++k^+{{p_\perp^2-2p_\perp k_\perp}
\over{k_\perp^2}}\right)^2}}
\end{eqnarray}
The transverse momentum integration in this expression is cutoff not
at the scale $k_\perp^2$ (which is of the order $Q^2$), but rather
$k_\perp^2/x$, which at small $x$ is a very large scale. Physically
the part of the integration region above $k_\perp^2$ corresponds to
emission of jets which are much harder than the probe. Precisely the
same problem is encountered in the standard perturbative
treatment~\cite{catani}.  These processes have to be considered
separately, and at this point we will disregard them.

Eq.(\ref{deltas}) looks very suggestive. Introducing the notation
\begin{eqnarray}
\delta \rho^a(x)= 2f_{abc}\alpha^{iN}_b(x) \delta A_{ic}^N(x)
\end{eqnarray}
we see that the linear coupling term between the soft field and the
hard fluctuation is of the form
\begin{eqnarray}
2{\mathrm{Tr}}\,\delta\rho(x) A^-(x)\delta(x^-)
\end{eqnarray} 

We would therefore like to integrate in the path integral over those
components of the fluctuation field $\delta A$, which are
``orthogonal'' to $\delta\rho$. In other words we would like to change
variables from $\delta A_{ib}$ to $\delta \rho^a$ and some $X^a$, and
integrate over $X$. In fact to get the result to the leading log
accuracy it is not necessary to do it explicitly.  Since $\delta \rho$
is linear in the fluctuation field, and the integral over the
fluctuation is Gaussian, it is clear that the result of the procedure
described above will be of the form
\begin{eqnarray}
\lefteqn{\int [d\,\delta\!A]\ \exp\{i\delta S\}} 
\\ & = & \nonumber
M(\alpha) \int [d\delta \rho]\exp\left\{-\int_{x,y}{1\over 2} 
\delta\rho^a(x) \delta\rho^b(y)
[\delta \chi]^{-1}_{ab}(x,y)+i\delta\rho^a(x) A^{-N}_a(x)\right\}
\end{eqnarray}
Here $M$ is the contribution of the determinant which arises in the
Gaussian integration over $X$. To the leading logarithmic accuracy
this contribution can be ignored.  This amounts to neglecting the loop
corrections with all particles in the loop having the longitudinal
momentum in the same range $P^+_N\le p^+\le P^+_{N-1}$. Corrections of
this type do not give large contributions at small $x$~\cite{catani}.
It was also shown in the previous analysis \cite{ayala}, that such
contribution could be ignored at small x for modifications to the
Weizs\"acker-Williams background field.

The matrix $\delta \chi_{ab}(x,y)$ is given by
\begin{eqnarray}
\delta \chi_{ab}(x,y)
& = & 4i\hspace{-2em}
\int\limits_{P^+_N\le|p^+|\le P^+_{N-1}} 
\hspace{-2em}
{dp^+\over 2\pi}
f^{acd}f^{bef}\alpha^c_i(x_\perp)\alpha^e_j(y_\perp)
 D^{Ndf}_{ij}(p^+, x_\perp,x^+,y_\perp,y^+)
\nonumber \\
\label{deltachi}
\end{eqnarray}

To proceed further, we need to know the structure of the propagator of
the hard fluctuations $D^N$.  Since the longitudinal momentum scale in
the propagator is large we can use a no recoil, or eikonal
approximation to incorporate the effect of interaction with the softer
fields $A^{N-}$. The calculation is given in Appendix B.
\begin{floatingfigure}{0.5\textwidth}
\epsfysize=4.5cm
\begin{center}
\begin{minipage}{4cm}
\epsfbox{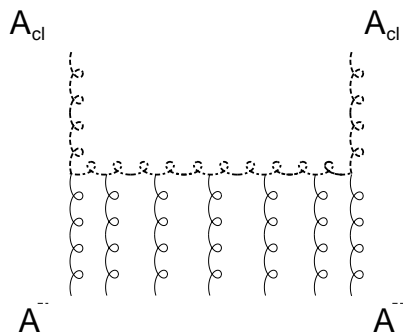}
\end{minipage}
\end{center}
\caption{\label{fig:softlegs} \small \itshape 
  Same as in Fig.\ref{fig:twolegs} but with additional insertions of
  the soft external field. These diagrams give corrections to the
  distribution function upon contracting the soft legs. They therefore
  correspond to the virtual corrections and are higher order in
  fluctuation fields.}
\end{floatingfigure}

In the eikonal limit, the propagator is
\begin{eqnarray}
\lefteqn{
D^{Nab}_{ij}(K^+,z^-;x^+,x_\perp,y^+,y_\perp)
} \\ 
& = & 
\delta_{ij}\delta (x_\perp - y_\perp)
{\mathrm{sign}K^+ \over K^+} i
\theta \left(\mathrm{sign}K^+(x^+ - y^+)\right)
\nonumber \\ && \times 
\hat{\mathsf P}
     \exp\left[ -i\int_{y^+}^{x^+} dz^+ 
        \mathbf{A}^-_{\mathrm{adj}}(z^+,z^-,x_\perp) \right]^{ab} 
\nonumber
\label{eikonal}
\end{eqnarray}
In this equation, $\mathbf{A}^-_{\mathrm{adj}}$ and hence the path
ordered phase is in the adjoint representation.  In the phase, the
dependence upon $x^-$ can be ignored since the function is slowly
varying on scales of the typical size corresponding to $1/k^+$.  Here
we have taken $K^+$ as the momentum conjugate to the difference of
coordinates $x^- - y^-$, and $z^- = (x^- + y^-)/2 \sim x^- \sim x^-$
is the average position associated with the field.

The expression above accounts only for soft fields with longitudinal
momenta smaller than that of the fluctuation $\delta A$.  In terms of
Feynman diagrams this corresponds to summation of the diagrams of the
type depicted on Fig.\ref{fig:softlegs}.  In fact, the propagator
$D^N$ also depends on the background field $\alpha^a_i(x_\perp)$ and
this dependence is important in parts of the phase space.  These
contributions are of the type Fig.\ref{fig:hardlegs}.  We will come
back to this point and discuss the importance of these terms later.
Temporarily, however we will disregard them in order to make the
discussion conceptually simpler.

As for the soft insertions, the following remark is in order. The
propagator depends only on the minus component of the vector
potential.  On the classical solution discussed in the previous
section, this component vanishes. It is therefore only the
fluctuations of $A^-$ around the new classical solution that
contribute to $D$. Since these effects are higher order in the
coupling constant, we will ignore them to this order. Again, these
corrections too are important at low transverse momentum.  This point
will be addressed in section \ref{sec:summary}.

With these approximations the fluctuation propagator becomes very
simple.  The fluctuation of the charge density $\delta\mu^2$ is time
($x^+$) independent and local in the transverse directions
\begin{eqnarray}
 \delta \chi_{ab}(x_\perp,y_\perp)=
{1\over {g^2\pi}}dy_N \delta^2(x_\perp-y_\perp)
\,f^{acd}f^{bed}\ \alpha^c_i(x_\perp)\alpha^e_i(x_\perp)
\label{dchi}
\end{eqnarray}

Note, that since our fields have a built in cutoff on the transverse
momentum, the $\alpha(x_\perp)\alpha(x_\perp)$ actually should be
understood as averaged on a transverse scale size $d^2x_\perp\sim
1/Q^2$.
 
We now make the approximation
\begin{eqnarray}
  \alpha^a_i(x_\perp)\alpha^b_j(x_\perp) \approx  
\langle\alpha^a_i(x_\perp)\alpha^b_j(x_\perp)\rangle=
{1\over 2(N_c^2-1)}\delta^{ab}\delta_{ij}\langle\alpha^2\rangle
 \label{one}
\end{eqnarray}

\begin{floatingfigure}{0.5\textwidth}
\epsfysize=4.5cm
\begin{center}
\begin{minipage}{4cm}
\epsfbox{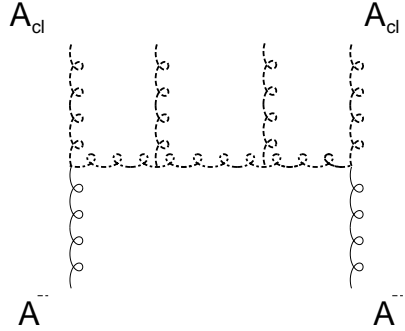}
\end{minipage}
\end{center}
\caption{\label{fig:hardlegs} \small \itshape 
  Same as in Fig.\ref{fig:twolegs} but with additional insertions of
  the hard background field. These diagrams are important in the
  region of momenta $Q$ of order $\alpha_s\chi$. See the discussion in
  the next section.}
\end{floatingfigure}

The averaging in Eq.~(\ref{one}) is over the distribution of $\rho$.
We believe this approximation should be adequate to describe the RG
flow of $\chi$, especially at large $Q$.  The fluctuations of $\rho$
are very short range in the transverse direction.  On the other hand
the fields $\alpha$ are slowly varying, its transverse correlation
length being of order $1/g^2\chi$. There is therefore very little
correlation between $\rho(x_\perp)$ and 
$\alpha(x_\perp)$ at the same point. Also, due to slow variation of
$\alpha(x_\perp)$ in space, approximation of $\alpha^2(x_\perp)$ by an
$x_\perp$ - independent constant should be good with accuracy
$g^2\chi/Q$.  Although this approximation is true on the average, it
ignores some of the correlations which are built into the longitudinal
structure.  It would be very important to study corrections to this
approximation or better yet to fully incorporate the structure of
Eq.~(\ref{dchi}) in the solution to the problem.  This is left for
further study.

We get therefore, that the change in the charge density is governed by
the parameter
\begin{eqnarray}
      \delta \chi_N(Q^2) & = & {1 \over {g^2\pi}}{N_c \over {(N_c^2-1)}}
 dy_N \langle\alpha_N^2\rangle_{Q^2}
\nonumber \\ & = &
{1 \over {g^2\pi}}{N_c \over {(N_c^2-1)}}
 dy_N \int_0^{Q^2}
 \hspace{-.5em}{d^2 k_\perp\over (2\pi)^2}\ 
 G^{aa}_{ii}(y, k_\perp)
\end{eqnarray}
Here $G^{aa}_{ii}(y, k_\perp)$ is the transverse Fourier transform of
the gauge field propagator Eq.~(\ref{propagator}).  The variation of
$\chi$ due to the change of the transverse cutoff is also easily
calculated
\begin{eqnarray}
  \delta \chi_N (Q^2) =  dQ^2 {N_c \over {(N_c^2-1)}} 
{1\over (2\pi)^2}{1 \over {g^2 \pi}}  \sum_{P=1}^N dy_P \ 
G^{aa}_{ii}(y, Q)
\end{eqnarray}

The change in the effective action is therefore
\begin{eqnarray}
     \exp\{i\delta S\} & = & \int [d\delta\rho_N]\  
\exp\left( -\int\hspace{-0.2em} d^2x_\perp  
{1 \over {\delta \chi_N(Q^2)}}{\mathrm{Tr}}\, 
\delta \rho^2_N(y,x_\perp) \right) 
\nonumber \\
& & \times\ \exp\left(
i \int d^2x_\perp  
\int\limits_{-\infty}^\infty\hspace{-0.2em} dx^+ \delta
\rho_N(x_\perp) 
A^-_N(x_\perp,x^+) \right)
\label{effact}
\end{eqnarray}

Now we identify some typical space-time rapidity for our source with
the momentum space rapidity.  We expect that $y_{\mathrm{space-time}}
\sim y_{\mathrm{mom}}$. Let us define
\begin{eqnarray}
  y = y_0 -\sum_{i=1}^N dy_i
\end{eqnarray}
where the the right hand side is the momentum space rapidity shifts
induced by integrating out the different scales.  We will see that the
left hand side has an interpretation of the space-time rapidity.  The
rapidity $y_0$ is the beam rapidity.  We see that the equation for the
evolution of $\chi$ generated by our renormalization group procedure
can be written as
\begin{eqnarray}
  d\chi(y,Q^2)/dydQ^2 = {N_c \over {N_c^2-1}}{1 \over (2\pi)^2}
{1 \over g^2\pi}\  G^{aa}_{ii}(y, Q)
   \label{group}
\end{eqnarray}
This equation can be either formulated as a BFKL type equation, when
one does the integral of $Q$ first, or as a DGLAP type equation if one
does the integral over $y$ first~\cite{QCD}.  Notice that it is a
non-linear generalization of both equations, since the right hand side
of the equation is a function of $\chi$.

This equation has a simple physical interpretation.  The factor of
$N_c/(N_c^2-1)$ is the charge squared per gluon.  The number of gluons
contained in our classical field is
\begin{eqnarray}
 {1\over S}{ dN\over dy dQ^2} = {1 \over (2\pi)^2} 
 {1 \over g^2\pi}\ G^{aa}_{ii}(y, Q)
\end{eqnarray}
where $S$ is the area of the nucleus.
What our analysis has shown is that the change in charge squared is
entirely due to the change in the number of gluons due to new phase
space opening up.  Of course this is a non-linear problem in general
since the source of charge changes the classical background field in a
non-trivial way.

Let us also define
\begin{eqnarray}
  \rho(y,x_\perp)  = \delta \rho_N(x_\perp) /dy
\end{eqnarray}
and 
\begin{eqnarray}
  A^-_N(x_\perp,x^+) = A^-(x_\perp,x^+,y)
\end{eqnarray}

Now we must compute the change in the classical field.  Since the
change in the classical field is small, we see that if we write
\begin{eqnarray}
  A_{N+1}^\mathrm{cl} = \delta A_N^\mathrm{cl} + A_{N}^\mathrm{cl}
\end{eqnarray}
we can linearize the equations for $\delta A_N^\mathrm{cl}$.  We find
that $\delta A^-_{N+1,cl} = 0$ and that
\begin{eqnarray}
  D^i(A_{N}^{cl}) \partial^+ \delta A_{Ni}^\mathrm{cl} + [\delta
  A_N^\mathrm{cl},\partial^+ A_{N}^{cl}] = g^2 \delta_N(x^-) \delta \rho_N /dy
\end{eqnarray}
In this equation, $\delta_N(x^-)$ means a delta function on the scale
of our new classical field, that is as a regularized distribution it
has its support on the scale $1/P^+_N$.  Now we define $\delta
A^\mathrm{cl}_N$ to vanish on the scale of the old classical field.
Therefore only the first term survives.

Upon identifying the index $N$ with the space time rapidity, we see
that our classical equation solves the equation we posited in the last
section that
\begin{eqnarray}
   D_i(A^{cl}) {d \over {dy}} A^i = g^2 \rho(y,x_\perp)  
\end{eqnarray}
(The easiest way to see this is to break up space-time rapidity into
discrete intervals.  Identify an index with each interval.  Precisely
the renormalization group equation for the field results.)

We also see that the path integral measure for the fluctuating field
is what we postulated in the previous section, with one caveat: We
have omitted some contributions which on the average vanish in their
contribution to the induced $\chi$.  These terms generate non-trivial
correlation in rapidity beyond which we compute.  We will not further
discuss their inclusion here except to note that they in principle are
computable, and should be included at some point.

\section{RG Equations for $\chi(y,Q^2)$}\label{sec:rgchi}

We now discuss the renormalization group Eq.~(\ref{group}).  This
equation determines how the color charge per unit area scales with
rapidity and a transverse resolution scale size $Q^2$.

The consistency of our analysis requires that the solution to the
renormalization group equation only involves information in the region
where our approximate methods of computation are valid.  It could
easily happen that the region of interest in transverse space after
several steps in the renormalization group procedure might drift to
some value where our approximations are no longer valid.  This might
happen if at some rapidity $y$, the relevant typical values of $Q^2$
became of order $\Lambda^2_\mathrm{QCD}$ or became much greater than
$\mu^2(y,Q^2)$ where our classical source size approximation breaks
down.  It is plausible that the region of integration for the solution
of the equations involves primarily the region of interest, since this
is physically where the field originates, but we have no proof.  In
addition, the region of large $Q^2$ where our classical methods no
longer apply is probably correctly treated even though the derivation
above breaks down.  In this region, the fields are weakly coupled, and
our expression derived by classical means appears to be correct even
in this region, to leading order in coupling.

The renormalization group equation may be formulated in the DGLAP form
by first integrating over $y$ as
\begin{eqnarray}
  d\chi/ dQ^2 = {{N_c } \over { (N_c^2-1)}} {1 \over (2\pi)^2} {1
    \over g^2\pi}
  \int^{y_0}_y\hspace{-.5em} dy^\prime\  G^{aa}_{ii}(y^\prime, Q)
\end{eqnarray}
It may be written in the BFKL-like form as
\begin{eqnarray}
  d\chi/ dy =  {{N_c } \over { (N_c^2-1)}} {1 \over (2\pi)^2} {1
    \over g^2\pi}
\int_0^{Q^2}\hspace{-.3em}
  dQ^{\prime2}\ G^{aa}_{ii}(y, Q^\prime)
\label{bfkllike}
\end{eqnarray}

Note that this is a nonlinear equation. The dependence of its right
hand side on $\chi$ is determined by the solution of classical
equations.

At this point we want to return to discussion of the terms that we
have neglected in deriving Eq.~(\ref{group}), namely the contributions
of the diagrams of Fig.\ref{fig:hardlegs}, with insertions of hard
background field. From the perturbative point of view those correspond
to modifications of the gluon distribution due to mixing between the
two particle and multi-particle (higher twist) operators.  These
diagrams can in principle be taken into account by using instead of
the free propagator in Eq.~(\ref{deltachi}) the full propagator in the
external field as calculated in~\cite{ayala}.

Although this calculation has yet to be performed, it is easy to
understand qualitatively the main modifications it will bring about.
First, even with the inclusion of the background field the additional
charge density $\delta\rho$ will remain static. This is due to the
fact that all the internal lines in the diagram Fig.\ref{fig:hardlegs}
have the frequency ($p^-$) of the order of the on-shell frequency
corresponding to the longitudinal momentum $P^+_N\le p^+\le
P^+_{N-1}$.  It is much smaller than the on-shell frequency of the
external line with momentum $k^+$. From the point of view of the
emitted particle, therefore the coupling is always to the static
source. The main effect of these extra insertions will be to modify
the right hand side of Eq.~(\ref{group}) by adding to it terms
nonlinear in $(\alpha)^2$.  This effect however will be significant
only for $Q$ of order $g^2\chi$.  Physically, the diagrams of
Fig.\ref{fig:hardlegs} describe an emission of the soft particle with
transverse momentum $k_\perp$ by a classical field $\alpha_i(x)$.
Clearly, as long as the transverse momentum of the emitted particle is
larger than the inverse correlation length of the field, the particle
is emitted locally. In this case the emission probability depends only
on $\alpha^2(x)$ at the point of emission. In the local limit
therefore the effect of these corrections will be of order
$\alpha^2(x)/k^2_\perp$.  At large $Q$ the main contribution to the
distribution function comes from large $k^2_\perp$, and the correction
due to nonlinearities is therefore negligible.  At $Q$ of order
$g^2\chi$ and smaller, the contribution of the diagrams in question is
important. However, in the saturation regime $Q\ll g^2\chi$ they do
not change the behavior qualitatively. In this region there is
practically no running of $\chi$ with $y$.  The reason is, that since
the correlation length of the classical field is of order
$(g^2\chi)^{-1}$, the phase space for emission shrinks to zero at
these values of momenta.

The qualitative features of the solution of our RG equation are these.
In the region of large $Q^2$, the equation approximately linearizes to
become
\begin{eqnarray}
  {d^2\chi\over dy\, d\!\ln Q^2} = 
    {{N_c \alpha_s } \over {\pi}} \chi
\end{eqnarray}
This is precisely the double logarithmic approximation to the DGLAP
equation~\cite{QCD}.  It would be solvable if it were not for the
dependence of $\alpha $ on $\chi$.  If we hold this fixed, we get
approximately (assuming that $\chi$ is a slowly varying function of
$y$ at some $Q_0^2$) that
\begin{eqnarray}
  \chi \sim \exp \left(2
    \sqrt{{N_c \alpha_s\over\pi}y\ln Q^2/Q^2_0}\right) 
\end{eqnarray}
The dependence of $\alpha_s$ upon $\chi$ will of course modify the
solution.

In the region of transverse momenta in the vicinity of the crossover
scale $\alpha_s\chi$ the nonlinearities in the renormalization group
equation become important.  It is likely, that one of the effects of
this will be that the transverse phase space in this region will be a
very slowly varying function of $Q^2$.  It is then more convenient to
turn to the form Eq.~(\ref{bfkllike}).  Assuming this to be the main
effect of the nonlinearities, and approximating the transverse phase
space by a constant $P$, we can write the solution as
\begin{eqnarray}
\chi=\chi_0\exp\left({N_c\alpha_s\over\pi}Py\right)
\end{eqnarray}
This has the BFKL - type behavior, growing as a power at small $x$.
To calculate the value of the constant $P$ we would have to include
virtual corrections which have been neglected so far.

Finally, in the saturation region where $Q^2 \ll \alpha_s^2
\chi(y,Q^2)$, the right hand side is constant up to logarithms.  Here
the solution is to a good approximation
\begin{eqnarray}
  \chi = \chi_0 + \kappa (y_0-y) Q^2
\end{eqnarray}
where $\kappa$ is some slowly varying function.  There is little
change until $(y_0-y) Q^2 /\chi_0$ becomes of order 1.

\section{Unitarity, Total Multiplicity and Summary}\label{sec:summary}

The issue of unitarity in deep inelastic scattering is related to the
$x$ dependence of
\begin{eqnarray}
  G(x,Q^2) = \int^{Q^2}_0 d^2p_\perp  {{dN} \over {dxd^2p_\perp}}
\end{eqnarray}
at fixed $Q^2$ as $x$ decreases.  We have seen that at fixed
$p_\perp$, there are two separate regions for $dN/dxd^2p_\perp$.  The
first is at large $p_\perp^2 \gg \alpha_s^2 \chi(x)$.  In this region,
the integral above is $xG(x,Q^2) \sim \ln(Q^2) \chi(x)$ up to factors
of logarithms of $\mu$.  As $x$ decreases, this is a rapidly rising
function of $1/x$.

At some point, for any fixed $Q^2$, the parameter $\chi $ will become
$\le Q^2$.  At this point, we are in the small $p_\perp$ region for
the computation of $dN/dxd^2p_\perp$.  In this region,
$dN/dxd^2p_\perp \sim \ln(p_\perp) $ up to factors of $\ln(\chi)$ (It
would be useful to determine these factors more accurately and
actually compute the cross section in this region, but again this is
beyond the scope of this paper.)  Here the structure function $xG$ has
at most a logarithmic dependence upon $x$.  There is therefore no
obvious contradiction with unitarity.  The dependence of $xG$ on $Q^2$
is also amusing, rising like some power of $Q^2$ up to logarithms,
until saturating at $\chi$.

The total multiplicity produced in hadron-hadron collisions at x may
also be estimated.  Here we return to rapidity variables.  On scaling
grounds alone, the multiplicity of produced gluons per unit area
should be $dN_g/dy\pi R^2 \sim \chi(y,\eta \sim 1)$.  These gluons
after production interact at high relative energy, and therefore
largely elastically.  The number of gluons should be approximately
conserved.  Later as a quark-gluon plasma is formed, the system
expands approximately isentropically, so that the total number of
gluons produced should be roughly the same as the number of pions.

At this point, we do not have a full solution of the renormalization
group equation in hand.  Suffice it to say that one expects rapid
growth of the parameter $\chi $ as $x$ decreases at large $p_\perp$.
This should be much faster than a power of a logarithm of beam energy.
The reason that this growth does not violate unitarity is because it
is arising from an enhanced contribution at larger transverse momenta.

The typical transverse momentum in this picture will go as the square
of the multiplicity per unit area.  The total deposited energy density
at a typical formation time $t \sim 1/\chi$ will be of order $\chi^4$.
All of these functions are expected to be asymptotically somewhat
rapidly rising functions of energy.

To summarize: The results of this work are extremely suggestive.  We
have presented a picture of low x gluon structure functions which has
many of the intuitive features normally associated with the pomeron.
The calculation presented here should be improved however in many
aspects.

We have made several drastic approximations in deriving the
renormalization group equation.  Let us once again point those out.

First, we have neglected the virtual corrections. Those are generated
by the diagrams on Fig.\ref{fig:softlegs} when one contracts the
external legs. Formally, as we mentioned in
section~\ref{sec:rgcharge}, those are higher order in fluctuation and
for that reason would seem to be sub-leading. However, some of these
diagrams are known to contribute to the BFKL equation, and therefore
must be important at least in some kinematic regime. This suggests
that the generic form of the effective action which we have relied on,
Eq.~(\ref{effact}) is not quite complete.  To see what is missing, let
us consider for a moment fields in three ranges of the longitudinal
momentum: the field $A_\mu(k^+)$, with $k^+\ge P^+_{N-1}$, the field
$B_\mu(l^+)$ with $P^+_{N-1}\ge l^+\ge P^+_N$ and the field
$C_\mu(m^+)$ with $m^+\le P^+_N$. The integration over $A$ and $B$
generates the effective action for $C$.  Taking into account the soft
insertions of Fig.\ref{fig:softlegs}, means we should use for the
fluctuations propagator the full eikonal expression
Eq.~(\ref{eikonal}) without setting the Wilson line factor equal to
unity.  The integration over the fluctuations of the field $A_\mu$
will generate the effective Lagrangian
\begin{eqnarray}
\lefteqn{i\int d^2x_\perp  
{\mathrm{Tr}}\, \delta \rho_{N-1}(x_\perp) 
{1 \over {\bar \chi_{N-1}(B^-+C^-)}}
\delta\rho_{N-1}(x_\perp)} 
\nonumber \\ && + 
\int d^2x_\perp  
\int\limits_{-\infty}^\infty\hspace{-0.2em} dx^+ \delta
\rho_{N-1}(x_\perp) \left[(B^-(x_\perp,x^+)+C^-(x_\perp,x^+)\right]
\end{eqnarray}
where
\begin{eqnarray}
\bar \chi_{N-1}(B^-+C^-)=\delta \chi_{N-1}W(B^-+C^-)
\end{eqnarray}
and
\begin{eqnarray}
W(A^-)=\hat{\mathsf P}
     \exp\left[ -i\int_{-\infty}^{\infty} dz^+ 
        \mathbf{A}^-_{\mathrm{adj}}(z^+,x_\perp, x^-=0) \right]
\end{eqnarray}
In the next step the integration over the fluctuations of $B$ leads to
the effective Lagrangian for $C$
\begin{eqnarray}
\lefteqn{\int d^2x_\perp  
\int\limits_{-\infty}^\infty\hspace{-0.2em} dx^+ \left[\delta
\rho_{N-1}(x_\perp) +\delta
\rho_{N}(x_\perp) \right]
C^-(x_\perp,x^+)} 
\nonumber \\ && 
+ i\int  d^2x_\perp  
{\mathrm{Tr}}\, \delta \rho_{N-1}(x_\perp) 
{1 \over {\langle\bar \chi_{N-1}(B^-+C^-)\rangle_B}}
\delta\rho_{N-1}(x_\perp) 
\nonumber \\ &&
+i\int d^2x_\perp  
{\mathrm{Tr}}\, \delta \rho_{N}(x_\perp) 
{1 \over {\bar \chi_{N}(C^-)}}
\delta\rho_{N}(x_\perp) 
\label{effactv}
\end{eqnarray}
In the second term the brackets denote averaging over the fluctuation
of the field $B$.
\begin{eqnarray}
\langle\bar \chi_{N-1}(B^-+C^-)\rangle_B=\delta \chi_{N-1}
\int [d\delta B] \ W(B+C)=(1-\gamma)\delta\chi_{N-1}W(C)
\end{eqnarray}
We see therefore, that the integration over the fluctuations in the
range of momenta between $P^+_{N-1}$ and $P^+_N$, not only generates
the additional charge density $\delta \rho_N$, but also modifies the
fluctuation amplitude of the charge density $\rho$ which is generated
by the higher momentum modes $k^+\ge P^+_{N-1}$.  This modification is
only important for the coupling of the density $\rho$ to the fields
with momenta $m^+\le P^+_N$, since only in this case the longitudinal
phase space is large and the correction factor $\gamma$ is proportional
to $\mathrm {log}(1/x)$.  In terms of Feynman diagrams this
calculation corresponds to the virtual corrections of
Fig.\ref{fig:virtual} and the factor $\gamma$ is directly related to
the so-called non Sudakov form factor~\cite{catani}.
\begin{floatingfigure}{0.5\textwidth}
\epsfysize=4.5cm
\begin{center}
\begin{minipage}{4cm}
\epsfbox{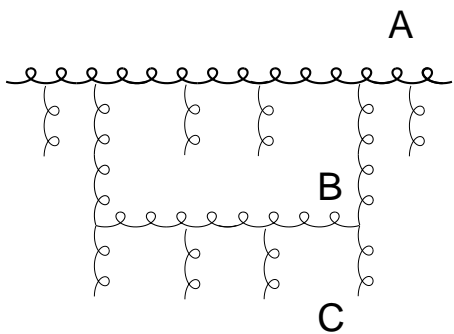}
\end{minipage}
\end{center}
\caption{\label{fig:virtual} \small \itshape 
  Virtual corrections of Eq.~(\ref{effactv}). The thickest line
  denotes the propagator of the field component $A$ with the largest
  longitudinal momentum. The thick line that completes the loop is the
  propagator of the field $B$. Finally, the thin external denote the
  external field $C$, for which the effective Lagrangian is being
  computed.}
\end{floatingfigure}
It is important to note, that even though the form of effective
Lagrangian Eq.~(\ref{effactv}) is not precisely the same as considered
in section~\ref{sec:extstruct}, on the classical solutions where
$A^-=0$ the two indeed coincide.  The solution considered in
section~\ref{sec:extstruct} is therefore still applicable to the
modified Lagrangian. The net effect of the virtual corrections is to
modify the running of the effective charge density $\chi(y)$ through
the change in the right hand side of the renormalization group
equation Eq.~(\ref{group}). This effect is calculable, and should
indeed be calculated, but this is beyond the scope of this paper. We
will only note, that since these virtual diagrams do not play any role
in the perturbative double log DGLAP treatment, and our
renormalization group equation reduces to it in the limit of large
$Q$, we expect these extra corrections to be important only in the
nonlinear regime $Q\sim\alpha_s\chi$.

As a second approximation we neglected the insertions of the hard
background field. This was discussed in the previous section, where we
have argued that these corrections are also unimportant at large $Q$.
Again, in principle these corrections are calculable if we employ the
full fluctuation propagator in the background field, as calculated
in~\cite{ayala}.

The third approximation in arriving to Eq.~(\ref{group}) was to
replace the square of the classical field by its average. This lead to
the absence of correlations in rapidity for the density fluctuations.
This approximation is also expected to be good for large transverse
momenta, where the emission of soft field is local in transverse
coordinates and therefore the additional density $\delta\rho$ is
practically uncorrelated with the density coming from higher
rapidities. At transverse momenta of order of the inverse correlation
function of the classical field this approximation should break down.

Clearly, the treatment of the nonlinear region $Q\le\alpha_s\chi$ in
the present paper is very rudimentary. It is in great need of
improvement, and we intend to address this problem in future work.  In
fact, at small transverse momenta the very notion of the momentum
cutoff $Q$ is highly questionable. The density - density correlation
which is given by the diagram Fig.\ref{fig:twolegs}, in momentum space
is a very slowly varying function of the transverse momentum at
$k_\perp\gg\alpha_s\chi$.  It depends on the momentum logarithmically.
In this range of momenta one can therefore approximate it by a
constant.  In our calculation precisely this is achieved by
introducing the cutoff $Q$ and using the eikonal approximation for the
propagator $D_N$, which results in a local correlator of density in
the transverse coordinates.  At momenta of order $\alpha_s\chi$,
however the correlator changes rapidly.  Approximating it by a
constant with some transverse cutoff should therefore result in an
error of order one. A more careful treatment will bring about
nontrivial transverse correlations of the charge density.  One should
therefore expect, that an improved treatment of low transverse momenta
region will modify the distribution for density fluctuations such that
nontrivial transverse as well as longitudinal correlations will
appear. The effective Lagrangian which would generate the classical
solutions will be of the form
\begin{eqnarray}
\int dy dy^\prime d^2x_\perp d^2 x_\perp^\prime
{\mathrm{Tr}}\, \rho(y,x_\perp)
\left[\mu^2(y,y^\prime, x_\perp,x_\perp^\prime)\right]^{-1}
\rho(y^\prime,x_\perp^\prime)
\end{eqnarray}
In fact, in a general case there is no reason to expect that the
weight will be Gaussian, so that the weight function $\mu$ could
itself depend on $\rho$.

It remains to be seen how large in fact will be the effect of these
improvements.  We believe, that although quantitatively it should be
significant, the qualitative picture presented here will be confirmed.
If the above issues can be answered in a satisfactory way, then one
can proceed to a detailed computation of hadronic interactions within
this approach.  Deep inelastic scattering from nuclei and Drell-Yan
production in heavy ion collisions might be computed.  The initial
conditions for nucleus-nucleus collisions might be found in detail,
The fluctuation spectra and correlations between fluctuations would
also do much to verify the above picture.  It would be very useful to
have data on the structure functions directly by using nuclei in HERA.

\section*{Acknowledgments}

We gratefully acknowledge the contributions of Rajiv Gavai and Raju
Venugopalan whose careful work showed that the original treatment
advocated by McLerran and Venugopalan led to infrared singular
correlation functions.  We thank Jianwei Qiu and Mark Strikman for
interesting discussions and Al Mueller for informing us of the work of
Yuri Kovchegov~\cite{yuri} which we received when this preprint was
being written.  This work was supported through DOE contracts DOE High
Energy DE-AC02-83ER40105 and DOE-Nuclear DE-FG02-87ER-40328. H.W. was
supported by the Alexander von Humboldt Foundation.

\appendix

\section{Normal Ordering the Distribution Function}
We have to calculate 
\begin{eqnarray}
    G_{ij}(y,x_\perp;y^\prime,x_\perp^\prime) & = & \int [d\rho]\   
   \exp\left( - \int dy^{\prime \prime}d^2x_\perp^{\prime \prime}
   {1 \over {2\mu^2(y^{\prime \prime},Q^2)}} 
   \rho^2(y^{\prime \prime},x_\perp^{\prime \prime}) \right) 
\nonumber \\ & \times &
   i^2 U(y,x_\perp) \nabla_i U^\dagger(y,x_\perp) \ 
   U(y^\prime,x_\perp^\prime) 
   \nabla_j U^\dagger(y^\prime,x_\perp^\prime)
\end{eqnarray}
This can be cast in the form
\begin{eqnarray}
\lefteqn{G_{ij}(y,x_\perp;y^\prime,x_\perp^\prime)}
 \\ & = & 
%
\langle \int\limits^\infty_y \hspace{-0.2em} dy^\prime 
     \left(U_{\infty,y^\prime} 
     \left( \nabla^i \Lambda(y^\prime 
     ) \right)
     U_{y^\prime,\infty}\right)(x_\perp) 
\int\limits^\infty_{\bar y}\hspace{-0.2em} d{\bar y}^\prime
     \left( U_{\infty,y^\prime} 
     \left( \nabla^i \Lambda(\bar y^\prime 
     ) \right)
     U_{\bar y^\prime,\infty}\right)(\bar x_\perp) 
\rangle_\Lambda
\nonumber
\end{eqnarray}
with the correlation function
\begin{eqnarray}
  \label{eq:lambdacorr}
  \langle \Lambda(y,x_\perp)\Lambda(y^\prime,x_\perp^\prime) 
  \rangle_\Lambda 
  & = & g^4 \mu^2(y,Q^2)\delta(y-y^\prime) 
        \gamma(x_\perp-x_\perp^\prime)
\end{eqnarray}
The function $\gamma$ is given in Eq.~(\ref{k4reg}) and as discussed
in section~\ref{sec:extstruct} is infrared singular. The leading
dependence on the infrared cutoff resides in the constant term
$\gamma(0)$. Fortunately, all terms containing $\gamma(0)$ cancel in
the expression for the correlation function.  To understand how these
cancelations work in (\ref{eq:lambdacorr}) let us first consider the
normal ordering of individual link operators first. To do so, let us
break up any link operator from $y$ to $\infty$ into ``infinitesimal
factors''
\begin{eqnarray}
  \label{eq:breaklink}
  U_{\infty,y}(x_\perp) & = & \lim\limits_{k\to\infty}
  \prod\limits_{n=1}^k
  U_{\infty,y_k}(x_\perp) U_{y_k,y_{k-1}}(x_\perp) 
  \cdot \ldots \cdot
  U_{y_1,y}(x_\perp) 
\end{eqnarray}
In the large $k$ limit each $U_{y_m,y_{m-1}}$ covers an infinitesimal
piece of the total path with a fixed length $\Delta = y_m-y_{m-1}$.
Due to the locality of (\ref{eq:lambdacorr}) in $y$ it is clear that
there will be no contractions between different factors in this
product. For an individual factor however we may expand and perform
the normal ordering
\begin{eqnarray}
  \label{eq:Uexp}
  U_{y_m,y_{m-1}}(x_\perp) & = & \id + i \int\limits^{y_m}_{y_{m-1}}
  \hspace{-.6em}dy\ \Lambda(y,x_\perp) +  i^2 \int\limits^{y_m}_{y_{m-1}}
  \hspace{-.6em} dy
  \int\limits^{y_m}_{y}
  dy^\prime\  \Lambda(y,x_\perp)\Lambda(y^\prime,x_\perp) 
  \nonumber \\ &&
  + O(\Delta^3)
  \nonumber \\ & = &
  : \id + i \int\limits^{y_m}_{y_{m-1}} \hspace{-.6em}dy\
  \Lambda(y,x_\perp) : + \id 
\left(i^2 \frac{g^4 N_c \gamma(0) }{2}\right) 
  \int\limits^{y_m}_{y_{m-1}}
  \hspace{-.6em}dy\ \mu^2(y,Q^2)
  \nonumber \\ && + O(\Delta^2)
\end{eqnarray}
where we have kept all terms up to order $\Delta$.  This is the only
non-suppressed tadpole contribution if the function $\mu^2(y)$ is
finite.  As a consequence, we have
\begin{eqnarray}
  \label{eq:prodU}
  \lefteqn{U_{y_m,y_{m-1}}(x_\perp)U_{y_{m-1},y_{m-2}}(x_\perp)}
\nonumber \\& = &
  : \id + i \int\limits^{y_m}_{y_{m-2}} \hspace{-.6em}dy\
  \Lambda(y,x_\perp) : + \id 
\left(-\frac{g^4 N_c \gamma(0) }{2}\right) 
  \int\limits^{y_m}_{y_{m-2}}
  \hspace{-.6em}dy\ \mu^2(y,Q^2)
  \nonumber \\ && + O(\Delta^2)
  \nonumber \\ & = &
  : U_{y_m,y_{m-2}}(x_\perp) : 
  \exp\left[\left(-\frac{g^4 N_c \gamma(0) }{2}\right)
  \int\limits^{y_m}_{y_{m-2}}
  \hspace{-.6em}dy\ \mu^2(y,Q^2)\right]
\end{eqnarray}
which immediately carries over to $U_{\infty,y}(x_\perp)$ upon
insertion into (\ref{eq:breaklink}).  The dangerous tadpole
contributions therefore can be factored out from a link operator by
writing it in the normal ordered form. Using this result we find
\begin{eqnarray}
\lefteqn{G_{ij}(y,x_\perp;y^\prime,x_\perp^\prime)}
 \\ & = & 
\int\limits^\infty_y \hspace{-0.2em} dy^\prime
\int\limits^\infty_{\bar y}\hspace{-0.2em} d{\bar y}^\prime \ 
     \langle 
     : \left(U_{\infty,y^\prime} 
     \left( \nabla^i \Lambda(y^\prime 
     ) \right)
     U_{y^\prime,\infty}\right)(x_\perp) : 
\nonumber \\ && \hspace{10em} \times
      : \left( U_{\infty,y^\prime} 
     \left( \nabla^i \Lambda(\bar y^\prime 
     ) \right)
     U_{\bar y^\prime,\infty}\right)(\bar x_\perp) :
\rangle_\Lambda
\nonumber \\ &&  \hspace{5em} \times
\exp\left[\left(-\frac{g^4 N_c \gamma(0) }{2}\right)\left(
  \int\limits^{\infty}_{y^\prime} + \int\limits^{\infty}_{\bar
    y^\prime}\right) 
  dy^\prime \ \mu^2(y^\prime,Q^2)\right]
\nonumber \\ & = &
 \int\limits^\infty_y \hspace{-0.2em} dy^\prime
 \int\limits^\infty_{\bar y}\hspace{-0.2em} d{\bar y}^\prime
  \ \tilde G_{ij}(y^\prime,x_\perp;\bar y^\prime,\bar x_\perp) 
\nonumber \\ &&  \hspace{5em} \times
\exp\left[\left(-\frac{g^4 N_c\gamma(0) }{2}\right)\left(
  \int\limits^{\infty}_{y^\prime} + \int\limits^{\infty}_{\bar
    y^\prime}\right) 
  dy^\prime \ \mu^2(y^\prime,Q^2)\right]
\nonumber 
\end{eqnarray}
The expectation value of the product of normal ordered fields we
dubbed $\tilde G$ does not contain any contractions within the
individual $U$'s. This is now easily evaluated order by order and then
resummed. Expanding the $U$'s to zeroth order, we have
\begin{eqnarray}
\tilde G_{ij}^{ab0}(y^\prime,x_\perp;\bar y^\prime,\bar x_\perp ) =  
    \delta^{ab}
      \delta(y^\prime-\bar y^\prime)\, g^4 \mu^2(y^\prime,Q^2) 
      \nabla_i \bar \nabla_j 
      \gamma(x_\perp-\bar x_\perp)
\end{eqnarray}

In the first order, a quick computation gives
\begin{eqnarray}
\lefteqn{\tilde G_{ij}^{ab1}(y^\prime,x_\perp;
\bar y^\prime,\bar x_\perp
  ) }
\\ & = & \nonumber
\tilde G_{ij}^{ab0}
       (y^\prime,x_\perp;\bar y^\prime,\bar x_\perp ) \ \
 (-g^4) (-N_c) \gamma(x_\perp-\bar x_\perp)
\int\limits_{y^\prime}^{y^0} \hspace{-.5em} 
  dy^{\prime\prime} \mu^2(y^{\prime\prime},Q^2) 
\end{eqnarray}
In this equation, $N_c$ is the number of colors.

Similarly, in $n'th$ order, we find
\begin{eqnarray}
\tilde G_{ij}^{abn}(y^\prime,x_\perp;\bar y^\prime,\bar x_\perp ) 
& = & 
     { ( g^4 )^n N_c^n \gamma(x_\perp -\bar x_\perp) \over n!}
\left[ \int\limits_{y^\prime}^\infty \hspace{-.5em} 
  dy^{\prime\prime} \mu^2(y^{\prime\prime},Q^2)\right]^{n} 
\nonumber \\ & &  \times \
  \tilde G_{ij}^{ab0}(y^\prime,x_\perp;\bar
  y^\prime,\bar x_\perp )  
\end{eqnarray}
Summing up, multiplying the tadpole factors and then performing the
remaining $y^\prime$ and $\bar y^\prime$ integrals now allows us to
write an explicit expression in which the leading infrared divergence
cancel. Assuming $y > \bar y$ the result is
\begin{eqnarray}
     G_{ij}^{ab}(y,x_\perp;y^\prime,x_\perp^\prime) & = &  
 -\delta^{ab}\left( \nabla_i \nabla_j^\prime 
\gamma(x_\perp-x_\perp^\prime) \right)
 {1 \over { N_c\left[ 
       \gamma (x_\perp-x^\prime_\perp)-\gamma(0)\right]}} 
\nonumber \\ && 
\left(1 - \exp\left\{g^4N_c \chi(y,Q^2) 
[\gamma(x_\perp-x^\prime_\perp)-\gamma(0)]\right\} \right)
\end{eqnarray}
where we have defined
\begin{eqnarray}
     \chi(y,Q^2) = \int_{y}^\infty
     dy^\prime \mu^2(y^\prime,Q^2)
\end{eqnarray}
The quantity $\chi(y,Q^2)$ is the total charge squared per unit area
at rapidity greater than the rapidity $y$.

\section{The eikonalized propagator }

In this appendix, we will derive an expression for the vector field
propagator in the eikonal approximation. We will solve the equation of
motion for the hard fluctuation field in the presence of external soft
vector potential. Throughout this analysis we neglect the effects of
the classical background field.

We start with the transverse component of the Yang-Mills equations:
\begin{eqnarray}
D_{\mu} F^{\mu i} = 0
\end{eqnarray}
and write the total field $A_{\mu}$ as
\begin{eqnarray}
A_{\mu} = \delta A_{\mu} + s_{\mu}
\nonumber
\end{eqnarray}
where $\delta A_{\mu}$ is the hard field describing the fluctuations
with high longitudinal momentum and $s_{\mu}$ is the soft field with
small longitudinal momenta only.

We assume that the only large momentum in the problem is the
longitudinal momentum of the hard field $\delta A_{i}$. Therefore, in
the equation of motion for the hard field we keep only those terms
which involve derivatives of the hard field with respect to $x^{-}$,
the coordinate conjugate to large momentum $p^{+}$. The equation of
motion for the hard field then becomes:
\begin{eqnarray}
\partial^{-} \partial_{-} \delta A_{i} - i [s^- , \partial_{-} \delta A_{i} ] = 
D^{-} (s) \partial_{-} \delta A_{i} = 0
\label{eq:eom}
\end{eqnarray}

To calculate the propagator we need to find eigenfunctions of the
operator $D^{-} (s) \partial_{-}$.  In order to do this we write
\begin{eqnarray}
\delta A_{i}^{\lambda} (x) = e^{ipx}\, \delta 
\tilde A^{\lambda}_{i} (x)
\nonumber
\end{eqnarray}
where the eigenvalue $\lambda =p^2$ and 
$\delta \tilde A_{i}$ is a slowly varying function of $x^{-}$. 
Then eigenvalue equation becomes
\begin{eqnarray}
D^{-}(s)\, \delta \tilde A_{i} (x) = 0
\nonumber
\end{eqnarray}
which has the solution
\begin{eqnarray}
\delta \tilde A^{a,\lambda}_{i,\alpha}\,(x,p) = 
\left[\hat{\mathsf{P}} \exp
(-i \int^{x^{+}}_{-\infty}\,
dz^{+} s^- (z^{+},x^- ,x_t)\right]_{ac} 
\epsilon_{i}^{(\lambda)}(p)\otimes 
u_{(\alpha)}^{c}
\end{eqnarray} 
where $a$ is the color label, $\lambda$ is the eigenvalue index,
$\epsilon_{i}^{(\lambda)}$ and $u_{(\alpha)}$ are the polarization
vector and color basis vector respectively.  The eigenfunctions
therefore are
\begin{eqnarray}
\delta A^{a,\lambda}_{i,\alpha}\,(x,p) =
 \left[ \hat{\mathsf{P}} \exp
(-i \int^{x^{+}}_{-\infty}\,
dz^{+} s^- (z^{+},x^- ,x_t)\right]_{ac} 
\epsilon_{i}^{(\lambda)}(p)\otimes 
u_{(\alpha)}^{c}\, e^{ipx}\, .
\end{eqnarray} 
The propagator is constructed as 
\begin{eqnarray}
G^{ab}_{ij} (x,y) = \int d\lambda \ 
\frac{\delta A_{i}^{a,\lambda} (x) 
\delta A^{\dagger b,\lambda}_{j} (y)}
 {\lambda -i\epsilon}
\nonumber
\end{eqnarray}
Since the soft field $s$ is a slowly varying function of $x^-$ , we
can neglect its variation with $x^-$ and write it as a function of
$z^-$ where $z^- \sim (x^- + y^- )$ is the average $(-)$ coordinate
associated with the soft field. The expression for the propagator can
be written as
\begin{eqnarray}
G^{ab}_{ij} (x,y) & = & \delta_{ij} \delta^{2}(x_t -y_t )
\int {{dp^+ dp^-}\over {(2\pi )^2}}
    {e^{-ip^+(x^- -y^- )}{e^{-ip^-(x^+ -y^+)}} 
    \over {-2p^+p^- -i\epsilon }}
\nonumber \\ && \times
 \left[ \hat{\mathsf{P}} \exp
(-i \int^{x^{+}}_{y^+}\,
dz^{+} s^- (z^{+},x^- ,x_t)\right]_{ab}
\label{eq:prop}
\end{eqnarray}            
Here we have used
\begin{eqnarray}
p^2 = -2p^+p^- + p^2_t \approx -2p^+p^-
\nonumber
\end{eqnarray}
and 
\begin{eqnarray}
\sum_{\lambda} \epsilon_{i}^{
(\lambda)}(p) \epsilon_{j}^{(\lambda)}(p) \otimes 
\sum_{\alpha} u^{c}_{(\alpha)} u^{d}_{(alpha)} = 
- \delta_{ij} \delta^{cd}.
\nonumber
\end{eqnarray}

The integration over $p^-$ is straightforward and gives a factor
proportional to $\theta (x^+ -y^+)$ for positive $p^+$.  To get the
propagator in momentum space, we Fourier transform with respect to
relative and center of mass coordinates, $x^- -y^-$ and $z^- \sim (x^-
+ y^-)$ to get:
\begin{eqnarray}
G^{ab}_{ij} (K^+, k^+, x^+,y^+,x_t,y_t) & = &
\frac{i}{2} \delta_{ij} \delta^{2} (x_t -y_t) 
\theta (x^+ -y^+) {{1}\over {K^+}}  
\\ && \nonumber \times 
 \left[ \hat{\mathsf{P}} \exp
(-i \int^{x^{+}}_{y^+}\,
dz^{+} s^- (z^{+},k^+ ,x_t)\right]_{ab}
\label{eq:fr}
\end{eqnarray}             
where $K^+$ and $k^+$ are the momenta conjugate to $(x^- -y^-)$ and
$(x^- +y^-)$ respectively. For negative $K^+$ we get the same
expression above with the argument of the theta function switched
around and a relative minus sign.

\end{document}